\documentclass[runningheads]{llncs}
\usepackage{graphicx}
\usepackage{pifont}
\usepackage{fancyhdr}
\usepackage{afterpage}

\begin{document}

\title{MORTAL: A Tool of Automatically Designing Relational Storage Schemas for Multi-model Data through Reinforcement Learning}
%
%
\author{Gongsheng Yuan\inst{1,2} \and
Jiaheng Lu\inst{1}}

%
%
\institute{University of Helsinki, FI-00014, Helsinki, Finland \\
\email{\{gongsheng.yuan,jiaheng.lu\}@helsinki.fi} \and
Renmin University of China, Beijing 100872, China}
\maketitle              

\thispagestyle{fancy}

\begin{abstract}

Considering relational databases having powerful capabilities in handling security, user authentication, query optimization, etc., several commercial and academic frameworks reuse relational databases to store and query semi-structured data (e.g., XML, JSON) or graph data (e.g., RDF, property graph). However, these works concentrate on managing one of the above data models with RDBMSs. That is, it does not exploit the underlying tools to automatically generate the relational schema for storing multi-model data. In this demonstration, we present a novel reinforcement learning-based tool called MORTAL. Specifically, given multi-model data containing different data models and a set of queries, it could automatically design a relational schema to store these data while having a great query performance. To demonstrate it clearly, we are centered around the following modules: generating initial state based on loaded multi-model data, influencing learning process by setting parameters, controlling generated relational schema through providing semantic constraints, improving the query performance of relational schema by specifying queries, and a highly interactive interface for showing query performance and storage consumption when users adjust the generated relational schema.

\end{abstract}

\section{Introduction}
The powerful transaction management ability, mature recovery mechanism, high availability, and excellent security of relational database management system (RDBMS) make it outstanding in the field of data management. Therefore, many commercial and academic frameworks reuse RDBMSs to store and query semi-structured data (e.g., XML, JSON) or graph data (e.g., RDF, property graph). For example, {\itshape MDF} (Mapping Definition Framework) \cite{Du2004ShreXMX} parses an XML schema based on the annotation method to get a relational schema and loads the XML document into tables.

\begin{figure}
\centering
\includegraphics[width=\linewidth]{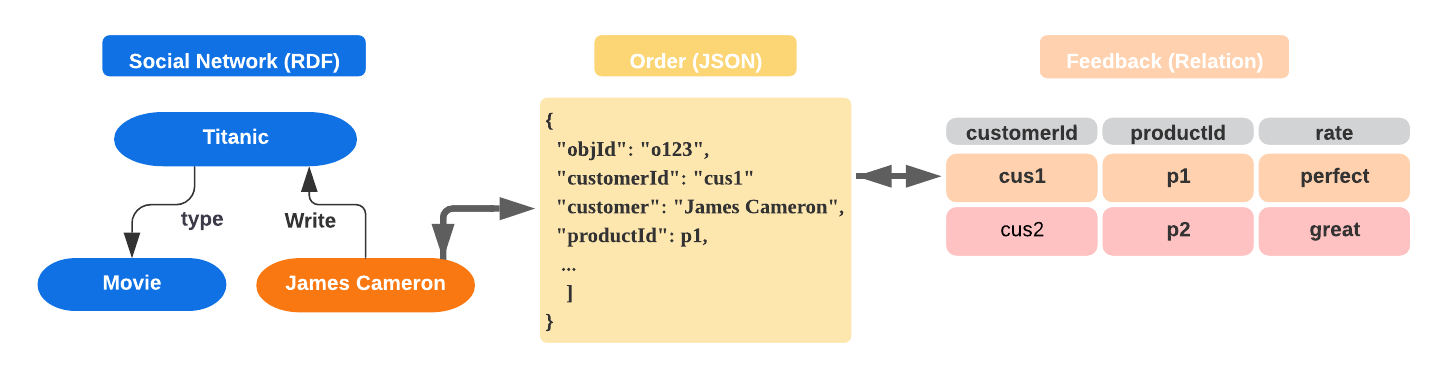}
\caption{An example of multi-model data.}
\label{fig:ERMutimodelData}
\end{figure}

The central task of existing works is to map a single data model into relational tuples. However, the swift growth of applications and devices diversifies data formats and makes managing these data in the same project difficult. Because it may lead to latency or data inconsistency when utilizing several databases in one project, such a dilemma calls for a novel multi-model database system. But it is expensive to develop such a novel system and replace the current popular RDBMS with it. Therefore, we consider mapping multi-model data (see Figure~\ref{fig:ERMutimodelData}) into relational data and use powerful RDBMSs to manage these shred data. Unfortunately, this idea is extremely challenging as it demands a great relational schema to store them while having an excellent query performance.

In this demonstration, we present a novel tool called MORTAL (transforming Multi-mOdel data into Relational TAbles based on reinforcement Learning) to store multi-model data in RDBMSs. Given multi-model data containing different data models and a set of queries, it could automatically design a relational schema to store these data while having a great query performance. 


As an important area of machine learning, reinforcement learning (RL) concentrates on how agents take actions in an environment to maximize the cumulative reward. In the standard RL model, it allows an agent to explore, interact with, and learn from the environment. On each step of interaction, the agent takes in the current state observations of the environment as the inputs and then chooses an action as the output. This action affects the environment by changing its state. Next, the environment produces a reward for that action and passes it to the agent. Then, the agent should choose actions that tend to maximize the long-run sum of rewards. This is achieved by systematic trial and error over time. The promise of a future high reward might lead to a non-best action for a certain iterative. RL achieves a trade-off between exploration (of unvisited area) and exploitation (of known knowledge) when interacting with the environment \cite{10.5555/1622737.1622748}.

Since our goal is to generate a relational schema having a great query performance (i.e., having minimum query time or maximum negative value of query time), it is similar to the goal of the RL model. Therefore, we use the RL model to address our problem. Specifically, we utilize Markov Decision Process to model the process of relational schema generation and let RL work with a dynamic environment to generate the optimal outcome.

\pagestyle{fancy}

\fancyhead{} 
\fancyfoot{}
\fancyhead[LE]{\scriptsize Gongsheng Yuan and Jiaheng Lu}
\fancyhead[RO]{\scriptsize MORTAL: A Tool of Automatically Designing Relational Storage Schemas}

%


\section{Design Philosophy}
{\itshape MORTAL} is designed to provide end-users with the convenience to easily get a great relational schema for storing any multi-model data source in RDBMSs while having a great query performance. Concretely, its design is based on the following two principles:

\begin{enumerate}
    \item \textbf{No need to extend RDBMSs.} Our RL-based approach should be able to work with any RDBMSs. With each interaction, our tool could obtain a relational schema by selecting an action. The final generated relational schema could be loaded into any RDBMSs without extending them.

    \item \textbf{Query-aware.} A key issue in generating relational schema to store multi-model data is that it could have a great query performance. Such a requirement naturally requires a query-aware approach. Fortunately, MORTAL (reinforcement learning model) satisfies this condition.
    
\end{enumerate}

\begin{figure}[h]
\centering
\includegraphics[width=\linewidth]{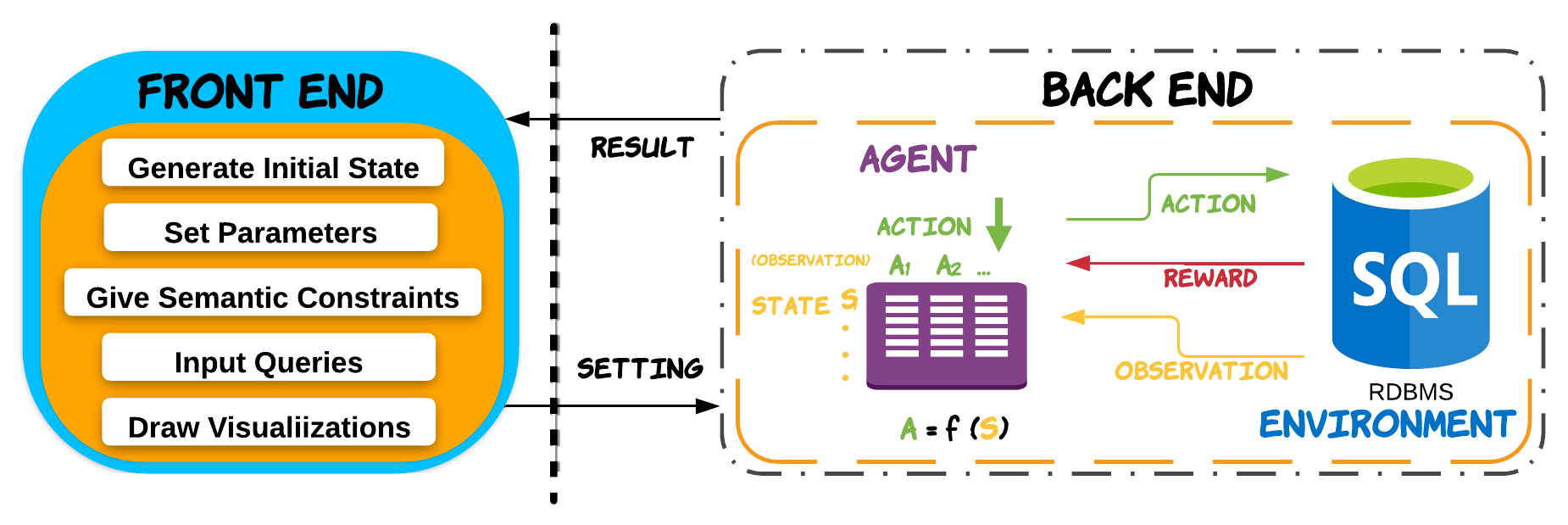}
\caption{Architecture of MORTAL.}
\label{fig:architectureofMORTAL}
\end{figure}

\section{System Overview}
Figure~\ref{fig:architectureofMORTAL} depicts the architecture of MORTAL. It consists of the following components. And \cite{gongsheng2021rl} gives more detail about this method.

\begin{figure}[t]
\centering
\includegraphics[height=10.5cm]{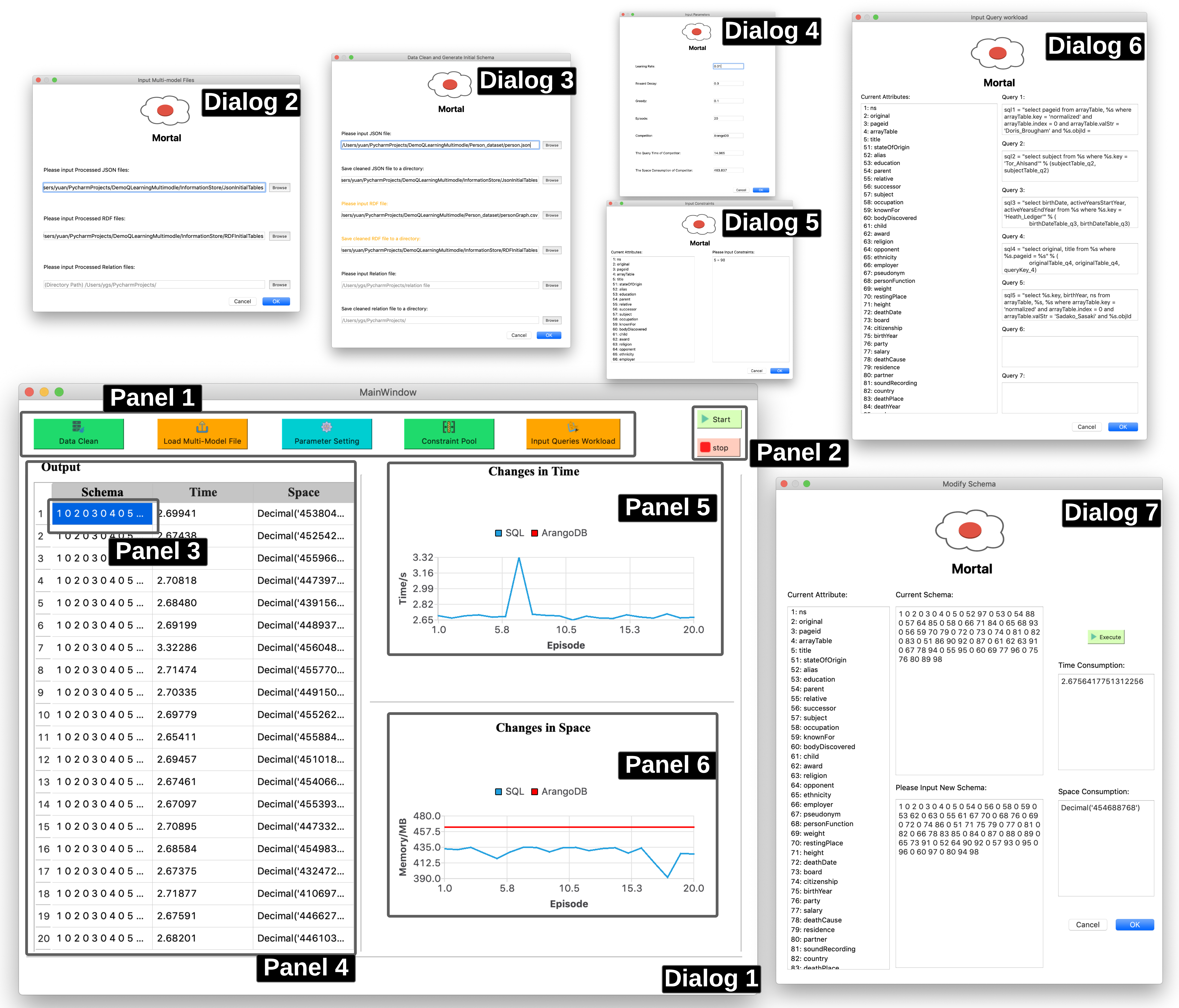}
\caption{The interface of MORTAL. (The icons on buttons are from ICONFINDER.)}
\label{fig:interface}
\end{figure}

\textbf{The GUI module.} Figure~\ref{fig:interface} depicts a screenshot of the MORTAL GUI. {\itshape Panel 1} contains a list of distinct buttons that enables a user to \ding{182} generate an initial relational schema and \ding{183} load it, \ding{184} set parameters (e.g., learning rate), \ding{185} specify semantic constraints, \ding{186} and input queries. After finishing setting {\itshape Panel 1}, users could use the buttons from {\itshape Panel 2} to start up learning or stop the whole program. {\itshape Panel 3} is like a hyperlink, which could open an interface for showing query performance and storage consumption after users adjusting selected relational schema. {\itshape MORTAL} uses {\itshape Panel 4} to manifest the generated relational schema in the process of learning, the query time over this relational schema, and space consumption of this schema in RDBMSs, and display the changes in time and space on {\itshape Panels 5} and {\itshape 6}, respectively. 

\textbf{Data clean module.} Since {\itshape MORTAL} needs to load data into RDBMSs to execute queries for getting the reward (the reduction of query time compared to the previous query), we would use a fully decomposed storage model (DSM) \cite{10.1145/971699.318923} and the model-based method \cite{florescu1999storing} to shred multi-model data into several little tables. Those little tables form the initial schema that is also the initial state of the RL model.

\textbf{Parameter module.} To make the RL model work, it needs to know the learning rate, reward decay, greedy, and the value of the episode. {\itshape MORTAL} adopts a variant Q-leaning called {\itshape Double Q-tables} as a learning algorithm to help choose actions. This method could reduce the dimension of the original Q-table \cite{watkins1989learning} and improve learning efficiency. With {\itshape Double Q-tables}, we define the action as a join operation. Therefore, the dimension of the Q-table is equal to the distinctive number of attributes (i.e., the number of little tables generated in the previous module). For each iteration, {\itshape MORTAL} firstly chooses one table (attribute) based on the first Q-table. Next, it selects another table (attribute) by the second Q-table to prepare to join. Based on the {\itshape Double Q-tables} method, users could try {\itshape MORTAL}'s best to explore different schemas by setting a small greedy value, or could assign a larger greedy value to accelerate converging of learning. Besides, the higher the given value of the episode, the higher probability {\itshape MORTAL} will find the optimal relational schema for the given queries and multi-model data. However, this would cost more time.

\textbf{Semantic constraints module.} There is one crucial issue in the progress of generating relational schema. That is, we need to take the relationships among the multi-model data into account. Besides, we previously introduce that {\itshape MORTAL} selects two tables (attributes) through two Q-tables for preparing the join operation in the \textbf{Parameter module}. But we do not know whether they could do the join operation for selected attributes. Here, the semantic constraints could help address this problem and tells the MORTAL what kinds of schemas users want to obtain.

\textbf{Interactive module.} Finally, this module allows users to input their own designed schema or adjusted relational schema selected on the {\itshape Panel 4} for observing or verifying its query performance and storage consumption.


\begin{figure}[t]
\centering
\includegraphics[height=6cm]{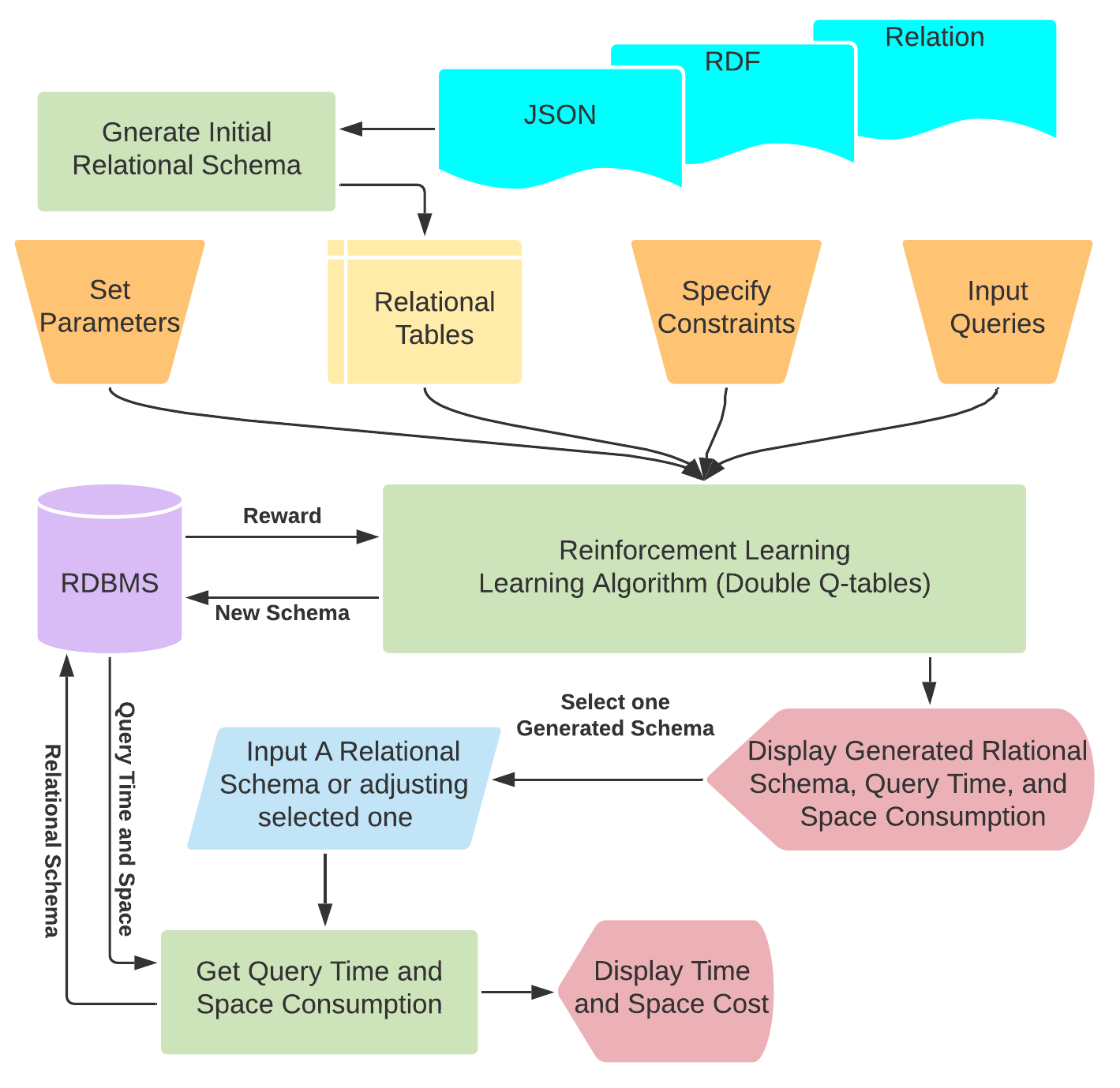}
\caption{The workflow of MORTAL.}
\label{fig:workflow}
\end{figure}

\section{Demonstration}
{\itshape MORTAL} is implemented with Python and PySide2. Our demonstration will be loaded with a multi-model dataset (Person) \footnote{https://www2.helsinki.fi/en/researchgroups/unified-database-management-systems-udbms/datasets/person-dataset} and show the results in Figure~\ref{fig:interface}. The key scenarios of the demonstration are as follows.

\textbf{Generate relational schema based on RL.} On the {\itshape MORTAL}'s main interface (see {\itshape Dialog 1} in Figure~\ref{fig:interface}), users firstly utilize the {\itshape Data Clean} button to open {\itshape Dialog 2} where users could get initial relational schema by multi-model data. Then users use the {\itshape Load Multi-Model File} button to feed the generated tables to {\itshape MORTAL} in {\itshape Dialog 3} to prepare to run the program. Next, with the {\itshape Parameter Setting} button, users could open {\itshape Dialog 4} to specify the value of parameters. In this dialog, users could also fill in the cost of query time and space for a competitor ( e.g., ArangoDB) to contrast that with {\itshape MORTAL}. Specifically, In Figure~\ref{fig:interface}, these values correspond to the red lines. For example, on {\itshape Panel 5}, since the query time of ArangoDB is far larger than that of {\itshape MORTAL}, so it does not show the red line. But on {\itshape Panel 6}, we could see that the red line is above the blue line. This is, the space consumption of relational schema is less than the ArangoDB's within these 20 episodes (set in {\itshape Dialog 4}) of learning. After that, users could use the {\itshape Constraint Pool} and {\itshape Input Queries Workload} buttons to open corresponding {\itshape Dialogs 5 and 6} and input what {\itshape MORTAL} needs according to the label's description. For example, users could provide a piece of information, ``5 = 98'' (see {\itshape Dialog 5} of Figure~\ref{fig:interface}), to let {\itshape MORTAL} know that the attribute ``title'' (``5'') in JSON is equal to the attribute ``title'' (``98'') in RDF, and they could be joined together. Finally, users could start up {\itshape MORTAL} by the {\itshape Start} button on {\itshape Panel 2}, and the results will be displayed on {\itshape Panel 3, Panel 5, and Panel 6}. Figure~\ref{fig:workflow} depicts this process.

\textbf{Interactive interface.} Using {\itshape Panel 3} or other items below it, users could open {\itshape Dialog 7} after double-clicks. In this dialog, users could input a new relational schema designed by users or adjust the selected relational schema to get its query performance and space consumption through pushing the {\itshape Execute} button. And users could sort the relational schema according to {\itshape time} or {\itshape space} by clicking the header of {\itshape Panel 4} to obtain the optimal schema.

\section*{ACKNOWLEDGEMENT}

The work is partially supported by the China Scholarship Council and the Academy of Finland project (No. 310321). We would also like to thank all the reviewers for their valuable comments and helpful suggestions.

%
%
\bibliographystyle{splncs04}
\bibliography{mybibliography}
%





\end{document}